# Vapor-solid-solid growth dynamics in GaAs nanowires

*Carina B. Maliakkal*[*,1,2,3], *Marcus Tornberg*[1,2,3], *Daniel Jacobsson*[1,3,4], *Sebastian Lehmann*[2,3], *and Kimberly A. Dick*[1,2,3]

[1]Centre for Analysis and Synthesis, Lund University, Box 124, 22100, Lund, Sweden.

[2]Solid State Physics, Lund University, Box 118, 22100, Lund, Sweden.

[3]NanoLund, Lund University, 22100, Lund, Sweden.

[4]National Center for High Resolution Electron Microscopy, Lund University, Box 124, 22100, Lund, Sweden.

*E-mail: carina_babu.maliakkal@chem.lu.se

**ABSTRACT**: Semiconductor nanowires are promising material systems for coming-of-age nanotechnology. The usage of the vapor–solid–solid (VSS) route, where the catalyst used for promoting axial growth of nanowire is a solid, offers certain advantages compared to the common vapor–liquid–solid (VLS) route (using liquid catalyst). The VSS growth of group-IV elemental nanowires have been investigated by other groups *in situ* during growth in a transmission electron microscope (TEM). Though it is known that compound nanowire growth has different dynamics compared to monoatomic semiconductors, the dynamics of VSS growth of compound nanowires has not been understood. Here we investigate VSS growth of compound nanowires by *in situ* microscopy, using Au-seeded GaAs as a model system. The growth kinetics and dynamics at the wire-catalyst interface by ledge-flow is studied and compared for liquid and solid catalysts at similar growth conditions. Here the temperature and thermal history of the system is manipulated to control the catalyst phase. In the first experiment discussed here we reduce the growth temperature in steps to solidify the initially liquid catalyst, and compare the dynamics between VLS and VSS growth observed at slightly different temperatures. In the second experiment we exploit thermal hysteresis of the system to obtain both VLS and VSS at the same temperature. The VSS growth rate is comparable or slightly slower than VLS growth. Unlike in the VLS case, during VSS growth we see several occasions where a new layer starts before the previous layer is completely grown, i.e. 'multilayer growth'. Understanding the VSS growth mode enables better control of nanowire properties by widening the range of usable nanowire growth parameters.

Controlling the electronic, mechanical and optical properties of semiconductor nanowires by tuning their crystal structure, composition and morphology enable their application in devices such as solar cells, electronics, light-emitting diodes, LASERS, photodetectors, thermoelectrics, biosensors and qubits.[1–4] Nanowires are most often grown using a foreign metal to seed the one-dimensional growth. The 'vapor-liquid-solid' (VLS) mechanism was proposed to explain this anisotropic growth[5] and was later confirmed through *in situ* observations.[6] According to the VLS mechanism, the atomic species constituting the semiconductor dissolve in the seed particle, forms a supersaturated metallic liquid alloy ('catalyst'), and subsequently precipitate to form the solid semiconductor.[7] An alternative vapor-solid-solid mechanism (VSS, in which the catalyst is solid instead of liquid) has also been proposed.[8,9] Some nanowire growths have been identified to occur with a solid catalyst on the basis of *ex situ* characterization of catalyst post-growth;[10–13] or by investigating the temperature range required to grow the wires[14] combined with equilibrium phase diagrams. However, effects such as size dependant decrease in melting point,[15–18] super-cooling[19–22] and thermal history[19] poses risks in an accurate assessment of catalyst phase in *ex situ* studies. Later on, *in situ* transmission electron microscopy (TEM)[19,23–26] and *in situ* reflection high-energy electron diffraction (RHEED)[27] studies have provided direct observation of VSS growth. It is thus very important to understand how the VSS process works and how it compares to the well-studied VLS growth process.

One major advantage of VSS growth compared to VLS is that in some material systems it can enable fast switching of materials in axial nanowire heterostructures.[23] During VLS growth the liquid catalyst acts as a reservoir of the nanowire species, and hence the switch from one composition to another is gradual.[28–30] On the other hand, the solubility of nanowire species in the solid catalyst particle is lower, enabling abrupt junctions, increased purity and better control.[23,27,31] Another advantage of VSS growth is that it greatly expands the range of possible catalyst materials, to include those with inappropriately high eutectic temperatures but potential advantages to for instance crystal structure control and incorporation of trace elements.[32] This would also enable the fabrication of nanowire-based devices in a way that is compatible with standard industrial processes.[33]

It is understood that nanowires mostly grow layer-by-layer along the nanowire-catalyst interface (for both VSS and VLS growths)[34,35]. The growth of each ledge, consisting of either one or more atomic layers, is often referred to as 'ledge-flow' (and sometimes also as 'step-flow'). A few groups have investigated the VSS growth of nanowires *in situ* in a transmission electron microscope (TEM),[19,23–26,36,37] a couple of which had a spatial resolution to directly measure the height of individual ledges[24,37]. Hofmann *et al.* observed during *in situ* VSS growth of Si nanowires that new ledge(s) can form even before the first ledge is completed, and that each ledge can be made of more than one atomic layer.[24] On the other hand, VSS Si-nanowire studies from another group reported that each ledge was only one bilayer thick (found indirectly due to limited spatial resolution)[25,26]. They occasionally observed steps of triple bi-layer height when the surface was unclean or for a small diameter nanowire that grew slowly even at higher precursor partial pressure.[37] On comparing VLS (Au-catalyzed) and VSS ($Cu_3Si$-catalyzed) Si nanowire growth happening at roughly the same growth rates (but different growth conditions), they observed that during VLS growth each bilayer grew rapidly once nucleated, but with long waiting time in between successive layer growths. On the contrary, in the VSS case the individual bilayers grew slowly but with short waiting times between successive layers.[25]

Note that the above mentioned studies on layer growth dynamics in VSS systems were on elemental semiconductors - either Si or Ge nanowires.[23–26] Although Si and Ge are very important semiconductor material systems (especially for electronics), as they are indirect bandgap materials they are inappropriate or less than ideal for certain optical applications. Many compound semiconductors, on the other hand, have a direct bandgap making them more efficient in such applications.[38] Hence it is very important to understand growth of compound nanowires. However, the layer growth dynamics in compound nanowires are different from monoatomic nanowires. For the VLS growth of a binary nanowire using a (foreign) metal seed, the two different component species could have very different miscibility in the metal catalyst, in turn affecting the kinetic processes.[39] Naturally this implies that the VSS growth of compound nanowires could be different from VSS growth of monoatomic nanowires. Therefore, it is important to separately study VSS growth of compound nanowires and compare the dynamics to the VLS route.

The aim of this study is to compare the growth dynamics of the VLS and VSS processes using the same catalyst material and similar growth conditions. Growth is performed in an aberration-corrected environmental TEM and observed *in situ* and *in operando*. We use two strategies in this study: (a) reducing growth temperature in steps and (b) cool the system followed by heating it to obtain solid and liquid catalysts at the same temperature, while in both cases keeping the precursor flow and the starting seed metal (Au) the same. The layer growth data studied for different temperatures for the first strategy is from one individual GaAs nanowire. The second strategy is employed using another single nanowire. We observe that the growth dynamics of GaAs are significantly different from the case of monoatomic VSS nanowire growth. We also observe that the growth rate during VSS growth need not necessarily be significantly different from VLS growth at similar conditions.

**Results and Discussion**

**Nanowire growth:** GaAs nanowires were grown *in situ* in an environmental TEM. TEM chips based on silicon nitride (amorphous) act as substrate and heater. The chips are engineered to have regions where the $SiN_x$ is etched out for improved spatial resolution. Au nanoparticles were deposited on the chips to seed the growth. The chips were heated to 420 °C and the precursor gases (trimethylgallium for Ga; arsine for As) were introduced to nucleate and grow the nanowires using MOCVD (metal organic chemical vapor deposition). At 420 °C these nanowires grow by the VLS mechanism. Two sets of experiments were conducted in this study, one where the temperature was decreased in steps to solidify the catalyst and the layer growth was studied at different temperature — spanning VSS and VLS growth. The second is an experiment where we initially recorded VLS growth at a particular temperature, then decrease temperature drastically to solidify the catalyst and then return to the earlier studied temperature to compare VLS and VSS growths at the same temperature.

**Stepwise cooling**

First let us discuss the experiment where the temperature was decreased in steps. The nanowire was nucleated at 420 °C and was growing by the VLS mechanism (Fig. 1 a) with a diameter of 22 nm. The growth temperature was then decreased in steps (Fig. 1 b), and at each temperature videos of the nanowire growth were recorded. At 310 °C the particle solidified (Fig. 1 c) while the growth continued. The observed lattice spacing (0.25 nm) is consistent with Au-Ga α' phase (~13% Ga). At temperatures between 420 °C and 330 °C the nanowire grew in the wurtzite structure and the catalyst-nanowire interface diameter did not change noticeably. At 320 °C and lower the wire grew in zinc blende structure. Note there was no change in crystal structure observed between VLS and VSS growth at similar temperatures (320 °C VLS and 310 °C VSS), and so this change to zinc blende at 320 °C is a consequence of temperature rather than catalyst phase.

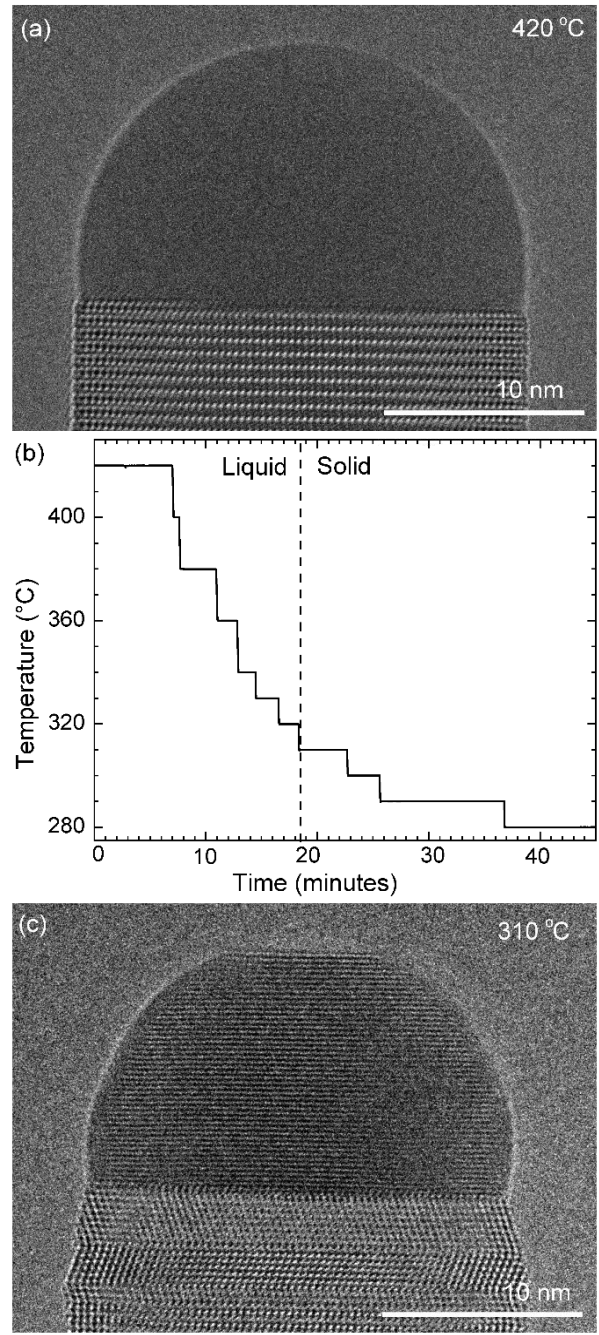

Fig. 1. **Stepwise Cooling**: (a) Bright-field TEM image of the nanowire- catalyst system during layer growth at 420 °C. At this temperature the catalyst is liquid. (b) The growth temperature was decreased in steps in this experiment. The catalyst solidified at 310 °C with one set of the crystal planes aligned parallel to the nanowire-catalyst interface. (c) TEM image where the catalyst has solidified.

We call the time each layer takes to complete since its nucleation as the 'layer completion' time (also sometimes called step-flow time in literature), and the difference between the ending of one layer and the start of the next layer as 'incubation time'. The incubation time and layer completion time were measured from the videos at each temperature and are plotted in Fig. 2 a. The x-axis of the plot is the layer number in ascending order but is discontinuous wherever the temperature is decreased. The incubation time before nucleation of each layer is plotted at the same x-value where its layer completion is positioned in Fig. 2 a. Sometimes a new layer starts to grow even before the previous layer has formed fully. When this occurs, it is denoted by a cross mark in the bottom panel of Fig. 2 a; this representation however does not give information about how far into the growth of the previous layer the new layer started. The information on how early or late during the growth of the previous layer(s) that the new layer starts can be visualised using Fig. 2 b-d. (In addition to data in Fig. 2 a, plots b-d also contains information of additional layers for which only part of the layer growth process was recorded, i.e. those at either the beginning or the end of each video.) The time we started recording videos at each temperature is set to 0s for that temperature in Fig. 2 b-d.

In the VLS mode the layers grew one at a time i.e. a new layer starts to grow only after the previous one has completely grown (Fig. 2 a, b); there were no occasions of double or multilayer growth. The time it takes for each individual layer to grow (layer completion) was very similar for 420, 380, 360, 340 and 330 °C where the catalyst was liquid, about 1.3 s. As the temperature is decreased from 420 to 320 °C, there is a very gradual increase of the incubation time (averaged value is plotted in Supporting Information Fig. S1). At 320 °C, where the crystal structure switches to zinc blende, the layer completion becomes faster. On further decreasing the temperature to 310 °C the first layer grew while the catalyst was still a liquid. At the start of the next layer (in the same video frame) the catalyst solidified with one set of lattice planes aligned parallel to the nanowire catalyst interface. The first two layers that nucleated after the catalyst solidified had shorter incubation time compared to the later layers (Fig. 2 a and Supporting Information Fig. S2). This can be attributed to a transient effect, as the particle initially has a higher supersaturation due to excess Ga just after solidification (lower Ga solubility in the solid phase than the liquid). Since the layer completion is expected to be limited by As availability[39] (due to the low solubility of As in the Au-Ga system[40]), the excess Ga is not expected to influence the layer completion time.

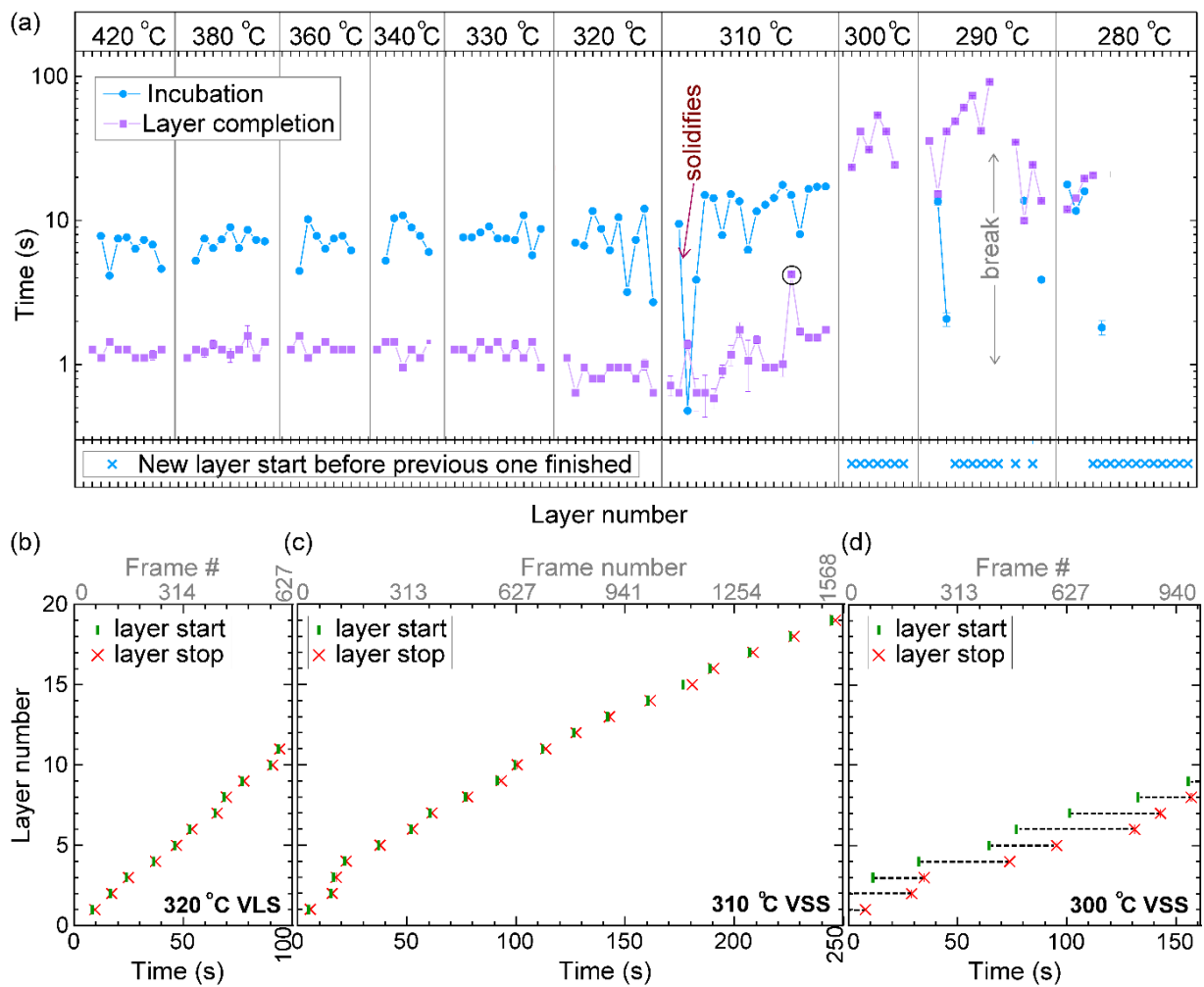

Fig. 2. **Layer growth at different temperature with solid and liquid catalyst**: (a) Incubation time and layer completion time growth at different temperatures in the top panel. The x-axis is the layer number plotted forward in time, but is discontinuous when the temperature is decreased. Each x-axis tick indicates one layer. Temperature is noted at the top. When a new layer starts before the previous layer finishes it is marked with a cross in the bottom panel. The starting and ending of each layer for successive layers are plotted for (b) VLS growth at 320 °C, (c) VSS growth at 310 °C, and (d) VSS growth at 300 °C. The x-axis shows the time from the start of each video. During 300 °C VSS growth new layers were seen to nucleate even before the previous layer is completely grown. In (b) - (d) the y-axis range is set to be the same.

The layer completion times for the solid catalyst at 310 °C, and liquid catalyst at 320 °C are similar, implying that the diffusion through the catalyst is not the rate limiting step here (because diffusion through solid phase is expected to be orders of magnitude slower than that through a liquid catalyst).[10] At 310 °C most of the growth occurred as single layers (each layer finished growing before the next one started). However, there was one instance of a double layer growth (indicated by the black ellipse in Fig. 2 a) that happened when the new layer(s) was twinned relative to the crystal orientation below (Fig. S2). This layer grew extremely slowly, suggesting that in case more than one layer is growing in parallel, then each of those layers would grow slowly. This is reasonable as the layer growth rate is proportional to the rate of arriving As atoms[39] and the time to complete each layer(s) naturally depends on the number of atoms required to form the layer(s). On decreasing the temperature further to 300, 290 and 280 °C, the catalyst remained in the solid phase. At these temperatures there were several instances of multiple layers growing simultaneously and each layer grew very slowly. The transition from a (mostly) single layer growth at 310 °C to a regime where more than one layer can grow simultaneously by a difference of just 10 °C indicates that the growth is very sensitive to different growth parameters. At 280 °C, multiple layers nucleated before the previous finished, followed by an evident displacement of the catalyst to the right side. Since this displacement interferes with accurate interpretation of the image data, the experiment was terminated at this point; only data for layers that started before this displacement is included in the plot (Fig. 2 a), and so completion time for these layers is not measured.

The diameter of the catalyst-nanowire interface did not change significantly while the wire was growing in the wurtzite structure (400 - 330 °C). However, when this nanowire was growing in the zinc blende structure, a change in the diameter of the nanowire-catalyst interface was observed as the growth progressed. In the VLS zinc blende growth (at 320 °C), the interface diameter decreased (from 21.8 nm to 21.2 nm). But in the VSS mode (zinc blende structure) the diameter kept increasing gradually and the catalyst height was subsequently decreasing at all investigated temperatures. Fig. 3 shows a few representative TEM images. Quantitative data on the change of interface diameter as a function of time is shown in Supporting Information Fig. S4. There was no apparent change in the lattice structure during VSS growth (refer Fig. 3 e, f). This thus changes the relevant dimensions of the system including GaAs-catalyst interface area, exposed surface area of the catalyst, the nanowire sidewall surface area and perhaps the catalyst volume – which in turn affects the amount of Ga and As required to form each layer, catalytic decomposition of the precursor species at the catalyst surface, the diffusion of the growth species in the catalyst (be it through the catalyst volume or through the catalyst-wire interface), collection of Ga adatoms from the sidewalls of the nanowire, etc. The system reshaping might be driven by the formation of low-energy facets. (Such drastic solid catalyst reshaping was observed in several experimentsfor relatively smaller Au seed (diameter <20 nm), while no significant reshaping was observed for 30 nm Au seeds.)

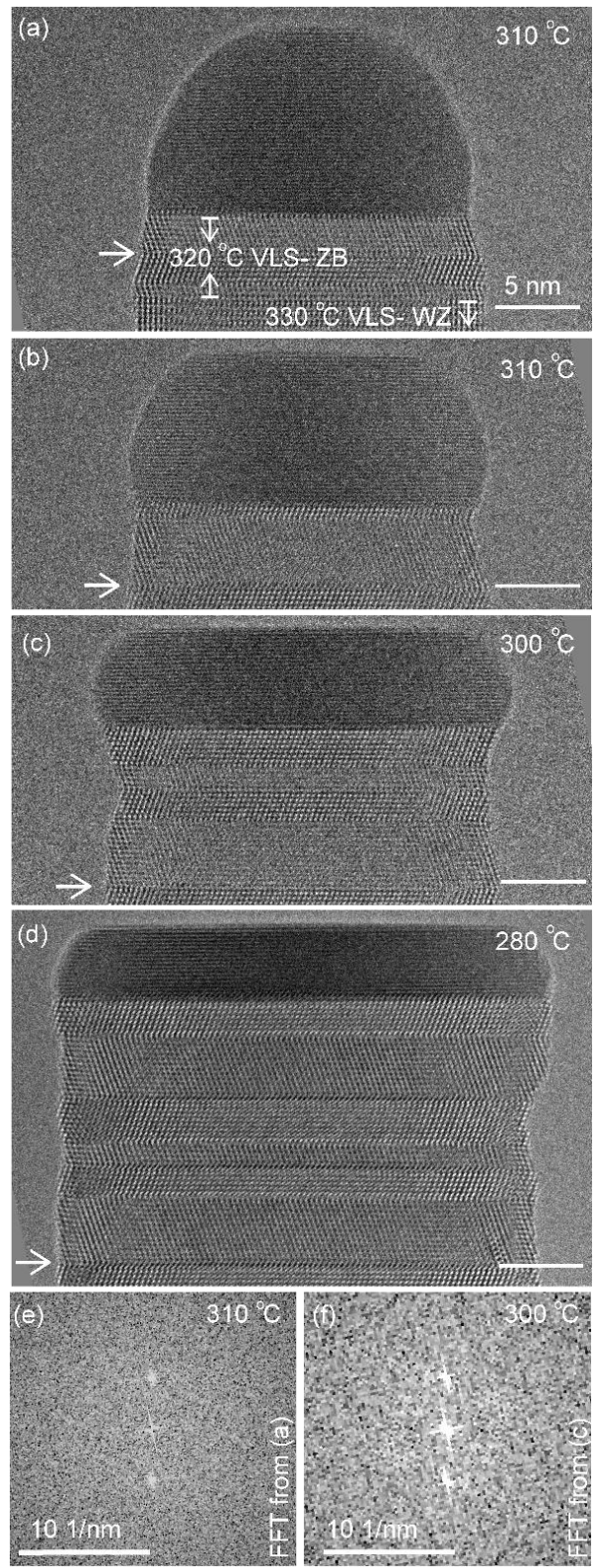

Fig. 3: (a-d) TEM images (after averaging and rotating) of the nanowire at the same magnification at different temperatures. The arrow marks the same defects in all the images. Scale bar is 5 nm in (a-d). Power Spectrum/Fast-Fourier-Transform (FFT) of catalyst region from images (a) and (c) are given as (e) and (f) respectively.

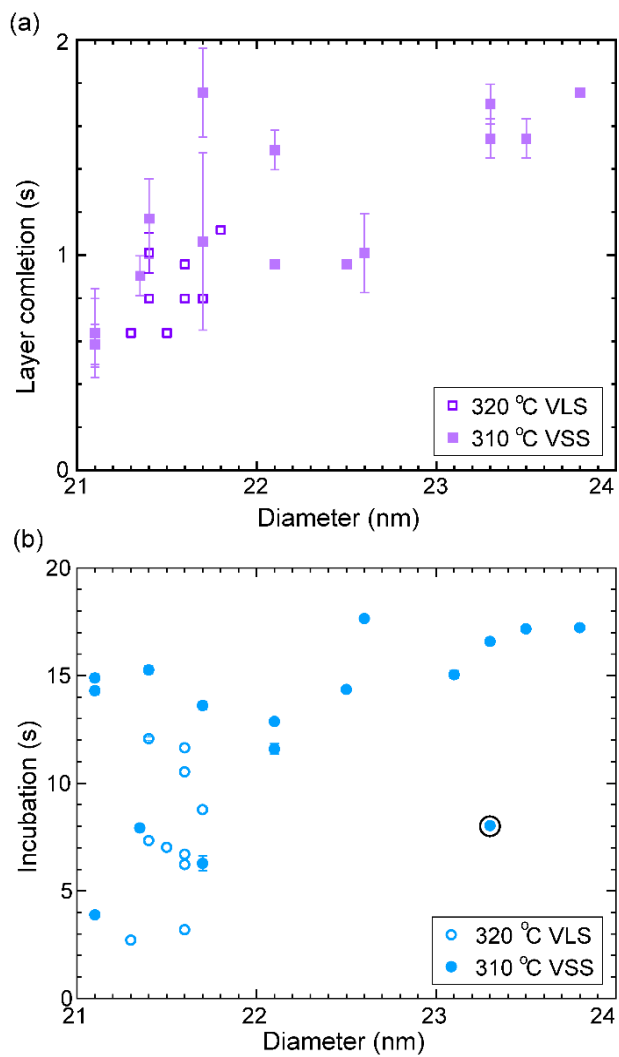

Fig. 4. **Growth times and diameter:** The layer completion time (a) and incubation time (b) are plotted against the nanowire diameter measured at the catalyst-nanowire interface. Data from two nearby temperatures where the catalyst phase changed is used here. Solid symbols denote layer growth times for the solid catalyst particle at 310 °C and the open symbols denote liquid catalyst at 320 °C. The range in plot (a) does not include the layer completion time of the twinned double layer (value 4.25±0.1 s) that grew. The incubation time right after this double layer was lower and in encircled in the plot (b). The error bars are related to the uncertainty in identifying the starting and stopping of each layer from the video.

Let us take a closer look at the layer growth times at the two temperatures across which the solidification occurred (i.e. 320 and 310 °C) in association with the diameter change. Fig. 4 shows layer completion time and incubation time as a function of the diameter of the wire-catalyst interface (y-axes are in linear scale). (The first couple of layers which showed a transient behaviour just after solidification at 310 °C are excluded from this plot.) We see that for similar diameter both the layer completion time and the incubation time are similar for the 320 °C VLS and 310 °C VSS growths. As discussed earlier, the diffusion through solid phases is typically orders of magnitude lower than through a liquid. Our observation of similar layer completion times for the solid and liquid catalyst phases implies that the diffusion to the growth front, irrespective of whatever the diffusion route can be, is not the rate limiting step.[10] With increasing diameter, we see that layer completion becomes slower. The layer completion is limited by the availability of As atoms in/at the catalyst.[39,41] Arsenic atoms reach the catalyst by direct impingement on the catalyst surface (rather than primarily by surface diffusion of As adatoms along the nanowire side facets to the catalyst).[42] One possible explanation for the increase of layer completion time with increasing diameter could be related to the increase of ratio between interface area to the exposed catalyst surface area (more details of this model considering geometric parameters are in Supporting Information section S4). Moreover, the nanowire width increases significantly at these conditions indicating that there could be competition for available growth material between the axial and radial growths. This will increase the incubation time, and perhaps increase the layer completion time too.

The average growth rate at the different temperatures studied are plotted in Fig. 5 (details of the measurement are given in Methods section). We can see that in the VLS mode there is a small decrease of growth rate with decreasing temperature, but there seems to be a relatively larger decrease of average growth rate at 310 °C when the growth mode switched to VSS (the comparison here is between (a) the observed VSS growth at 310 °C and (b) a linearly extrapolated value for hypothetical VLS growth at 310 °C based on the VLS growth at higher temperatures). Some earlier reports studied monoatomic nanowire growth and reported one or two orders of magnitude slower growth rate for VSS compared to VLS.[19] However, in this study of GaAs compound nanowires the observed difference between VLS and VSS nanowire growth rate is not as drastic. Moreover, note that the catalyst reshaping and nanowire widening will also influence the average growth rate. As discussed earlier, the layer completion time and the incubation time for the 320 °C VLS and 310 °C VSS growths are very similar (Fig. 4) for similar nanowire diameter. This in turn implies that the growth rate (which is essentially inverse of sum of incubation and layer completion times) are also very comparable between 320 °C VLS and 310 °C VSS for comparable dimensions. Thus, the difference between average growth rate in the VLS and VSS case observed here is at least partly due to the change in morphology.

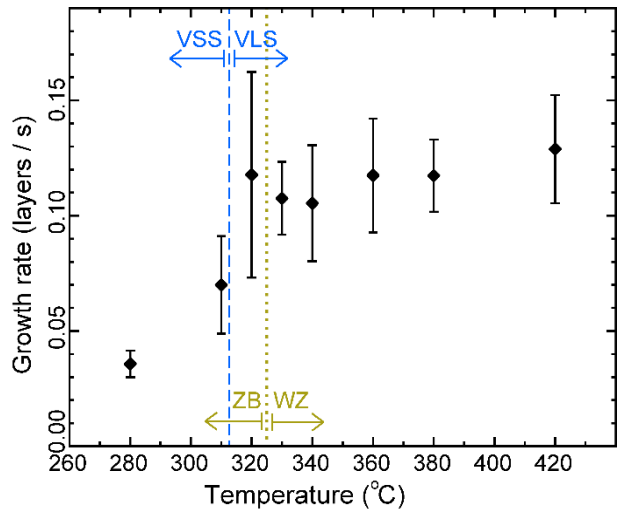

Fig. 5. **Average growth rate**: Average growth rate of the same nanowire at different temperatures. At 310 °C and lower the catalyst is solid. The wire was growing in the wurtzite (WZ) structure at temperatures between 420 °C and 330 °C and as zinc blende (ZB) at lower temperatures. Error bars represent standard deviation.

In this experiment the VLS and VSS growth are compared at different temperatures. Temperature can change different parameters like (gas phase) thermal decomposition of precursors, Ga adatom diffusion on NW sidewalls, Ga and As solubility in the catalyst, As evaporation rate, flow patterns in the growth cell and surface energies. It is also possible to compare the VLS and VSS at the same temperature by exploiting thermal hysteresis. This approach also has a limitation— that it relies on a transient state and not a steady state phenomenon. Hence it is important to use both these approaches and compare the results. The second approach of thermal hysteresis is what we discuss next.

**Experiment utilizing thermal hysteresis**

The phase of the catalyst is a function of not just the temperature but also the thermal history or hysteresis.[19] In the second experiment we use this to obtain VSS and VLS growth at the same temperature. Nucleation of the nanowires was performed at 420 °C, just like the previously discussed experiment. The diameter of this nanowire was 27 nm. After nucleation the growth temperature was decreased directly to 280 °C without pausing at intermediate temperatures to allow the system to reach a steady state (Fig. 6 a); this procedure enabled the catalyst to remain a supercooled liquid at this lower temperature. We set time = 0 when the sample temperature reached 280 °C (on decreasing from 420 °C) and use this time reference in Fig. 6. After observing some layer growth events at 280 °C the temperature was further decreased in steps as shown in Fig. 6 a, to allow the particle to solidify.

From this low temperature where the particle was a solid, the temperature was increased to 260 °C and then again to 280 °C and the VSS growth was monitored. Representative parts of video recording showing VLS and VSS growths at 280 °C are given as Supplementary videos. The observed lattice spacing (0.22 nm) in the solid catalyst is again consistent with Au-Ga α' phase (~13% Ga), similar to the earlier discussed stepwise cooling series. In this experiment the diameter changes were rather small (± 1-2 nm) and we did not see any significant catalyst reshaping as in the first experiment. In that stepwise cooling series with a smaller Au seed size perhaps there was some effect of the high surface-to-volume ratio effecting the energy balance. Other probable explanations for not seeing catalyst reshaping in the hysteresis series could be related to earlier formation of favorable facets or that the rearrangement of the solid was hindered by the lower temperature at which solidification occurred.

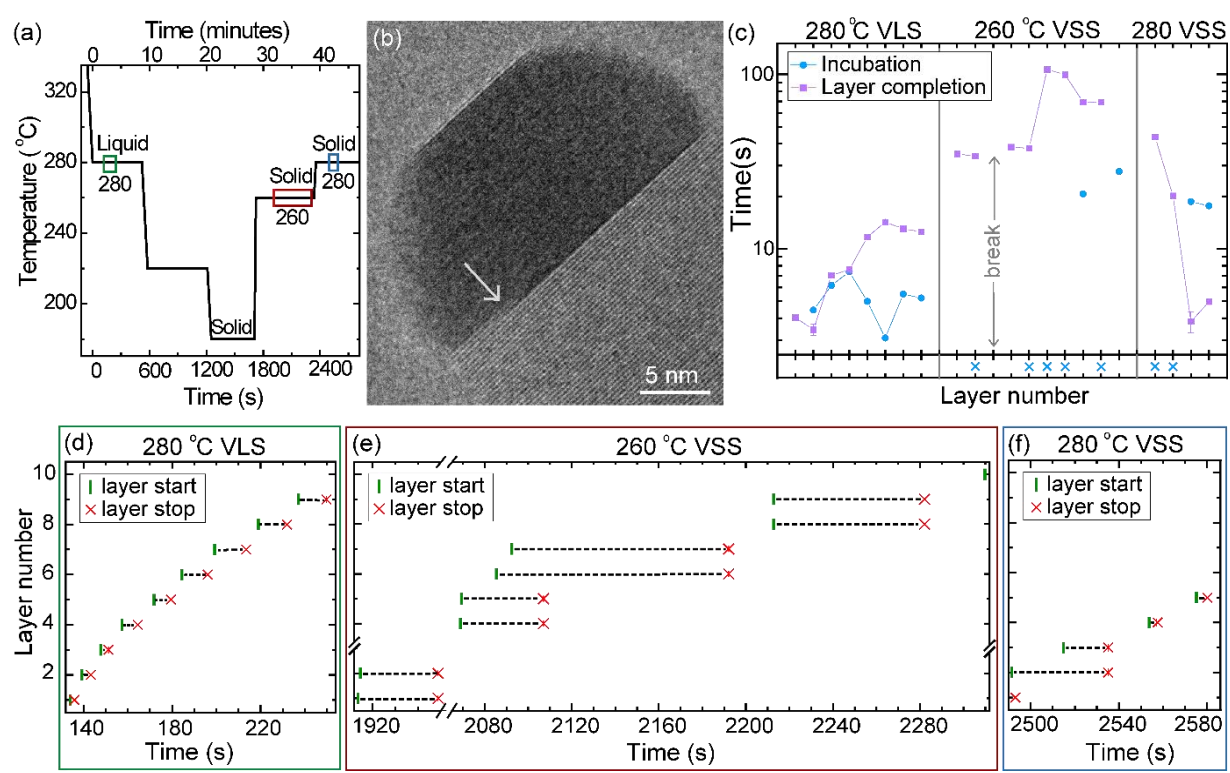

Fig. 6. **Thermal hysteresis experiment:** (a) Sample temperature as a function of time. The three colored boxes correspond to the time interval used for data in sections (d) – (f). (b) TEM image of the nanowire-catalyst system during VSS growth at 280 °C. In this TEM micrograph we see two layers growing simultaneously (marked by the arrow). (c) Incubation time and layer completion time for each layer during VLS growth at 280 °C and VSS growth at both 260 °C and 280 °C. The x-axis is discontinuous at the temperature changes. Each tick on the x-axis indicates one layer. The starting time (green triangle) and ending time (red square) of each layer is denoted for successive layers during the (d) VLS growth at 280 °C, (e) VSS growth at 260 °C and (f) VSS growth at 280 °C. We see in the VSS cases that often a new layer nucleates while the previous layer is still growing. In (e) we can also see occasions where a double layer has nucleated simultaneously. In (b) - (d) the x- and y-axis scales are set to be the same across panels.

The layer completion times and incubation times for VLS growth at 280 °C, VSS growth at 260 °C and VSS growth at 280 °C are shown in Fig. 6 c. (As for the previous experiment, when a new layer starts to grow even before the previous layer has formed fully the incubation time is denoted by a blue crossin Fig. 6 c.) During VSS growth at 260 °C there are instances where double layers have nucleated simultaneously. The layer completion time at 280 °C VLS and VSS are comparable for single layers, although double layer growth occurs more slowly as previously observed. The incubation time during VSS growth is higher than the VLS here at 280 °C. At a slightly lower temperature (260 °C) while the catalyst was a solid there were several instances were more than one layer was growing. We can also see an instance where 4 layers were growing in parallel (Fig. 6 e).

**Discussion**

**Comparison with monoatomic nanowire studies:** In monoatomic VLS systems the layer completion was observed to be effectively instantaneous while the incubation was very long.[23,25] Material required to supersaturate the system was sufficient to form one complete layer instantly once nucleated in that case. On the contrary, in monoatomic VSS cases the incubation time became very short and the layer completion time became long.[23,25] The solubility of the nanowire species in the solid catalyst particle was low and thus a small excess was enough to supersaturate the catalyst and start a new layer.[23,25] (This low solubility in solid catalyst is often the case, though not true universally for all systems.) Such a drastic contrast of incubation and layer completion times between VLS and VSS reported in monoatomic nanowires is not observed in this study of compound (GaAs) nanowires.

Even for the VLS route, the layer growth dynamics in compound nanowires are significantly different from monoatomic nanowires.[39] In the VLS growth of binary nanowires (GaAs studied as an example), the two elements forming the nanowire dissolve in the metal catalyst very differently.[40] Indeed it is common for binary semiconductors that one of the constituents alloys more easily with the seed catalyst, while the other constituent could have a much lower solubility. Specifically, it is common that the metallic element (group II and III elements such as Ga) alloys readily with frequently used seed metal (e.g. Au), while the non-metallic element (most group V or VI elements, such as As) have poor miscibility in the catalyst.[43] Under typical growth conditions this implies that the nucleation of a new layer is triggered by the chemical potential of the more miscible species, but the layer completion of each layer is restricted by less miscible species due to the scarcity of available atoms.[43] Layer completion in compound nanowires even for VLS growth is not instantaneous;[39] unlike the instantaneous layer completion in VLS monoatomic nanowire growth case[25]. It is thus not surprising to see even qualitative differences between compound and monoatomic nanowires for the VSS growth mode as well. Unlike seen in earlier reports on monoatomic NWs[23,25], the VSS incubation time observed here for GaAs in the single layer growth regime is not necessarily shorter than VLS case, especially when accounting for changes in nanowire diameter. Incubation time on the other hand is typically a bit longer for VSS GaAs than for VLS (in cases where layers grow one at a time).

**Growth rate difference between VLS and VSS:** Let us next discuss about the overall growth rate of VSS growth of nanowires in comparison to the VLS growth. Often in literature VSS growths are assumed to be slower than VLS, sometimes by one or two orders of magnitude. In their report on *in situ* Au-catalyzed Ge nanowire growth Kodambaka *et al.* mentioned that the VSS growth rate was one or two orders of magnitude slower (when temperature and precursor pressure were kept the same, while the thermal hysteresis was exploited to obtain VLS and VSS data at the same temperature).[19] However there are also reports with a much smaller difference. Chou *et al.* reported VSS growth of Si using Au-Ag catalysts to be a couple of time slower than VLS (when experiment was also done exploiting thermal history).[31] Comparable growth rates for VLS and VSS has been reported on the basis of *ex situ* growth studies also.[44,45] In our study of GaAs nanowire growth, the VSS growth seems to be slightly slower than VLS growth for comparable growth conditions. In summary, how different the VSS growth rate is in comparison to VLS depends on the specific chemical system, growth parameters and limiting processes (including precursor decomposition, diffusion etc.).

**Multiple layers growing simultaneously in GaAs VSS:**

*In situ* TEM investigations reported on VLS growth observed layer growing only one after the other[23,25,46] (except in one case of GaN nanowire growth[47] which we be briefly addressed below). On the other hand, for the VSS growth of monoatomic nanowires, multiple layers growing simultaneously has been observed in the case of Si nanowires[24–26] and was predicted to happen at low 'step mobility'[25]. With the terminology used in this report this means long layer completion time. In the GaAs experiments studied here we never observed more than one layer growing at any time during VLS growth. But in the VSS case we frequently observe that a new layer starts to grow even before the earlier layer finished forming. We also observe that in the VSS case nucleation of a double layer is also possible (i.e. we see two layers starting 'simultaneously' within the video resolution.) Gamalski et al. showed that double layers could grow together if the line energy of a ledge with two layers is appropriately low.[47] We observe occasions where in spite of long layer completion times a new layer has not started (see first few layers at 280 °C in Fig. 2 a). This could be either due to nucleation being a stochastic (or probabilistic) phenomenon, or that the slow step mobility cannot be the sole reason for multilayer growth. To explain how multiple layers can grow in parallel we propose a heuristic model based on energetics, material availability, and most importantly different solubility in the liquid versus solid catalyst.

In the VLS case, nucleation of each layer occurs only after overcoming a significant nucleation barrier; since the total number of atoms in the catalyst is small, the formation of a nucleus reduces the supersaturation sufficiently to prevent the formation of a new nucleation before the layer has finished growing. In the case of a solid catalyst, the solubility of Ga and As in the catalyst is very low. Thus, any extra Ga or As increases chemical potential significantly, seemingly much higher than the nucleation barrier. In such a scenario, the nucleation barrier becomes effectively less important and the requirement of a certain 'critical' supersaturation is not a major bottleneck. The rate limiting step is thus dominated by the sticking coefficient (or attachment rate) and material availability. This enables the starting of a new layer at the periphery due to the material supply directly from the vapor and additionally because it acts similar to a 'defect' or 'kink' site. The attachment process is still more favorable at the already growing layer than as a fresh layer at the rim of the catalyst-nanowire interface (we know this since we do not see several multilayers only just starting at the sides without growing into full layers). However, as the nucleation barrier becomes negligible for VSS, attachment of just a few atoms at the periphery can be sufficient to start a new layer, and there is a finite probability of this occurring before the already-growing layer is complete. This makes multilayer growth relatively more probable in VSS than in VLS. A possible explanation for the VSS growth at 310 °C showing a layer after layer growth behaviour could be the higher temperature compared to the other VSS temperatures studied here – as temperature increases the effect of each individual Ga atom on the chemical potential decreases, increasing the nucleation barrier and the potential for the growth to occur in a nucleation-limited regime. Among all the reports of VLS growth of different systems studied by *in situ* TEM, the report of multiple ledges growing simultaneously is by Gamalski *et al.*[47] on GaN nanowires growth; we suspect that behavior was due to the extremely low solubility of N in even the liquid Au-Ga catalyst which makes the situation very similar to the low solubility VSS case we discuss here for GaAs, in turn enabling multilayer growth.

**Summary**

Vapor-solid-solid growth of GaAs nanowires is compared with the vapor-liquid-solid growth by *in situ* investigation in a TEM. The phase of the catalyst particle is a function of not just the temperature but also of thermal history, so two distinct experimental strategies were designed to separately study each of these two effects. The VSS growth rate was found to be slightly slower than VLS, but comparable, rather than substantially slower as observed in some other systems. Earlier studies on monoatomic nanowires reported VSS growth to have shorter incubation and longer layer completion compared to VLS growth. Here, unlike the monoatomic studies, we see that the incubation time for VSS growth of Au-seeded GaAs nanowires is not necessarily very different from the VLS counterpart. Layer completion time was found to be very similar for VLS and VSS at comparable growth condition, indicating that the diffusion of reactants through the solid catalyst particle during VSS growth is not the growth rate limiting step. Moreover, observed differences in layer completion time and to some degree the incubation time could be largely accounted for by considering changes in the catalyst geometry. We also observed that while VLS usually proceeded as single layers, in VSS growth there is a high chance of more than one layer growing simultaneously. In VSS growth there can also be two layers nucleating simultaneously. How different the growth rate would be between VLS and VSS depends on thermal history, growth conditions, how the shape of the catalyst and nanowire evolves, the material system etc. Hence, when any particular aspect of VSS growth is advantageous for growing nanowires for a specific technological application, for example for the formation of sharp compositional heterostructures, one should not discard the idea suspecting slow growth, but should instead experiment it.

**Methods**

**Instrumentation:** Hitachi HF-3300S environmental transmission electron microscope (ETEM) with a cold field emission gun and a CEOS B-COR-aberration-corrector was used for growing nanowires *in situ.* The growth was performed in an open cell configuration on $SiN_x$-based MEMS chips from Norcada. The Blaze software supplied by Hitachi was used to control the sample temperature in a 'constant resistance mode' where the effects of gas pressure and material deposition on the temperature were compensated by a feedback mechanism. The heating chips used had windows of electron-transparent $SiN_x$ and also holes patterned on. The nanowire growths reported here were performed while the nanowire had grown into these hole regions and thus there is no $SiN_x$ at the background of the images shown here.

**Precursor supply:** The ETEM was connected to a gas handling system with the MOCVD gases. Trimethylgallium (TMGa) and arsine ($AsH_3$) are the precursors used here. Gas flows were controlled by mass flow controllers and pressure valves, and monitored during growth with a residual gas analyzer in the exhaust gas which had been calibrated to give partial pressures at the sample. $AsH_3$ was supplied directly without any dilution. $H_2$ was bubbled through the TMGa bubblers maintained at low temperature (bubbler bath temperature was -20 °C for the stepwise cooling series and -10 °C for the thermal hysteresis experiment). The TMGa/$H_2$ mixture was further diluted with hydrogen and a fraction of it was flown to the ETEM.

***In situ* growth:** GaAs nanowires were grown with Au as seed particles. TMGa and $AsH_3$ are introduced at 420 °C. The TMGa flux was briefly increased to trigger nucleation. Once nucleated the TMGa flow was stopped and we searched for nanowires which were growing towards the hole in the $SiN_x$ and are appropriately aligned – i.e. the nanowire-catalyst interface was parallel to the electron beam direction. Due to the relatively small tilting range available with these holders, it was not always possible to align the nanowire to a zone axis. However, from the lattice spacing along the growth direction the wires were found to grow along the ⟨111⟩ direction (while growing zinc blende) or the equivalent ⟨0001⟩ direction (while growing wurtzite). Each 'layer' we refer to is a 'GaAs bilayer' i.e. consists of one plane of Ga atoms and one layer of As atoms [layer forming the (111) plane in the case of zinc blende and (0001) for wurtzite]. Very often, while growing at these low temperatures, the catalyst particle topples to the side of the nanowire or the nanowire kinks. The two nanowire cases reported here did not kink during the experiment.

**Data acquisition and measurements**: Blaze software developed by Hitachi is for heating the sample and also logs the temperature. The precursor fluxes that flownto the ETEM were monitored with a residual gas analyzer (SRS RGA 300) using mass spectrometry. The precursor flows were calibrated to find the partial pressure at the sample. The TMGa partial pressure values reported here are estimated using this calibration and the mass spectrometry values measured. Further details of the experimental setup can be found in previous publications.[40,48]

The layer growth dynamics is recorded as videos made up of a series of TEM images. The growth of each layer is identified as a dynamic change in the contrast at the interface. In the TEM image we only observe a projection of the 2-dimensional plane. Some times in the projection it appears to grow from one side to the other. There are also instances where we see a contrast change start at somewhere in the middle (in the projected image). Occasionally there are instances were a contrast difference is seen along majority of the interface and just increases is strength- this could be when a layer is growing front to back (or back to front). The start or end of a layer is not always evident; in such conditions we analyze the video over and over to identify it and note down the values. The layer completion times and incubation times of individual layers are found from the starting and ending times of layers. The inaccuracy in measurement of the starting and ending of individual layers leads to the error bars in Fig. 2 a y-axis, Fig. 2 b-d x-axis, Fig. 4 y-axis, Fig. 6 c y-axis and Fig. 6 d-f x-axis. Naturally when the error bar is smaller than the size of the symbol used it is not visible in the figure. In Fig. 2 a and Fig. 6 c, when a new layer starts to grow before the previous layer has completely grown the incubation time is denoted by a blue cross (x).

**Average growth rate:** The average growth rate shown in Fig. 5 is calculated as the number or layers grown in an interval of time, where effectively the starting and ending of the time interval is not in between the growth of a layer. Mathematically it is the reciprocal of sum of average incubation time and average layer completion time; where only the incubation and layer completion time during single layer growth is used. At 290 °C and 300 °C there was always at least one layer growing, which is why no average growth rate is given in the plot for these two temperatures. At 280 °C there is only two measurements of incubation time and layer completion time, making it the average growth rate a very crude estimate.

**Specifics about stepwise cooling series experiment**
Temperature ramp rate for all the steps were 3 °C/s. The SiN$_x$ heating chips bulges due to thermal expansion as a function of temperature. When the temperature is changed the sample height making the sample out of focus. The video is not acquired when the temperature is changed and a new video is started after adjusting the sample height. This effect is lesser at lower temperatures – so the delay between reaching the temperature and starting the video (or analysis) & starting video is typically smaller. So, some layer growth events were not recorded causing the x-axis in Fig. 2 a to be discontinuous. In this specific experiment the time difference between reaching the new temperature and starting the video is as follows:  420 °C (have been at this temperature for very long), 380 °C : 103 s, 360 °C : 30 s, 340 °C : 20 s, 330 °C : 16 s, 320 °C : 9 s, 310 °C : -2 s i.e. we had started recording 2 s before temperature was reached, but there were no layers growing in that 2 s), 300 °C : 12 s, 290 °C :51 s and 280 °C : 2 s. The TEM images in Fig. 3 correspond to time 18 minutes 33 s in (a), 20 minutes 10 s in (b), 25 minutes 6 s in (c), and 36 minutes 52 s in (d), where time is measured as in Fig. 1 b.

Au aerosol particles of nominal sizes ~15 nm and ~20 nm were deposited on the $SiN_x$ chips prior to this growth experiment. During the stepwise cooling series, the precursor pressures near the sample were indirectly estimated to be T series TMGa=$6 \times 10^{-5}$ Pa of TMGa and 0.6 Pa of AsH3. A custom-made double-tilt holder from Hitachi High-Technology, Canada was used for the stepwise cooling series. The precursor gases were supplied in this case using stainless steel injectors opening in the microscope pole piece gap. The TEM video (bright-field) were acquired using Gatan OneView IS camera an exposure time of 0.159593 s for each image frame, giving 6.27 frames per s.

**Specifics about thermal hysteresis experiment**

As mentioned previously, after nucleating the nanowires the temperature was initially decreased and then increased to compare VLS and VSS at the same temperature. Temperature ramps were done with microscope gun valve closed during this experiment. The ramp rate was 1, 5 and 1 °C/s for the initial 280 °C (VLS), 260 °C and the final 280 °C (VSS) respectively. The time delay between reaching the new temperature and starting the recording was 100 s, 191 s and 134 s respectively. At the beginning of the first 280 °C (VLS) growth, an atomically thin ordered surface was present at the left and right sides of the catalyst-vapor interface. A couple of layers grew with this ordered surface but these layers are not included in Fig. 6. In principle, the particle would have solidified if we had maintained the system at 280 °C for an extended period of time. However, since at these low temperatures nanowires were very prone to kinking or toppling of the catalyst particle, waiting till the particle solidifies on its own was a risky option.

Au aerosol particles of nominal sizes ~30 nm were deposited on the $SiN_x$ chips prior to this growth experiment. For this experiment, a single-tilt holder with two separate microtubes running within the holder so release the precursors very close to the sample was used. The holder and the gas-handling system are connected by a polymer-coated thin quartz tube (PEEKSil) from Trajan Scientific. Precursor pressures were TMGa=$1.5 \times 10^{-4}$ Pa, As= 1.1 Pa. Video was recorded using an AMT XR401 sCMOS camera at a frame rate of 18 fps.

**Acknowledgements**

We thank Krishna Kumar, Robin Sjökvist and Michael Seifner for technical assistance during some experiments and later discussions. We acknowledge financial support from the Knut and Alice Wallenberg Foundation, NanoLund, and the Swedish Research Council.

# Supporting Information:
# Vapor-solid-solid growth dynamics in GaAs nanowires

**S1) Average incubation and layer completion time for the stepwise cooling series**

The Fig. 2 a of the main article showed incubation and layer completion time individual layer growth events. Fig. S1 shown here is the averaged value at each temperature. The first few layers grown at 310 °C (i.e. temperature where catalyst solidified) are not included in this averaging (the first layer at 310 °C had grown while the catalyst was liquid and the first two layers soon after catalyst solidification showed a reduced incubation time due to the transition).

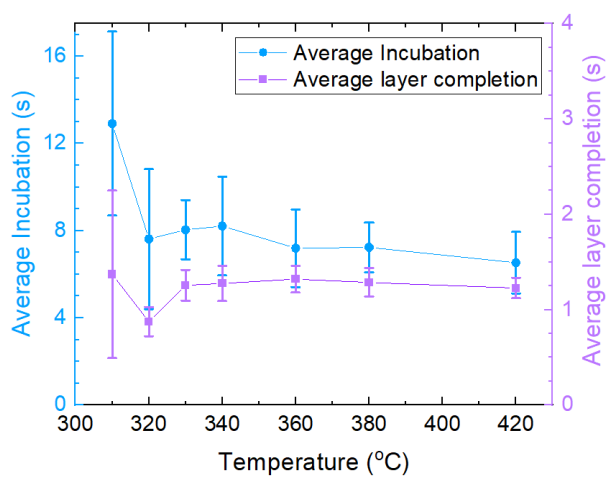

Fig. S1. Average incubation and layer completion time at each growth temperature for the stepwise cooling series. Y-axis are in linear scale.

**2) Individual layer growth events at 310 °C alone in stepwise cooling series**

The catalyst was initially a liquid at this temperature. The first layer at 310 °C grew while the catalyst was still a liquid. After a few seconds, the catalyst solidified and a new layer nucleated at the same time (within the 0.16 s temporal resolution of the video). The image where we observed solidified in Fig. S3 b. The image just before is given in Fig. S3 a. (These images show the raw data, without any image rotation.) The first two layers that nucleated soon after solidification showed a reduced incubation time due to the transition. We observed two occasions of the formation of twinned zinc blende layer. The first was a single bilayer and showed a slightly higher value of layer completion time than the other layers (marked with a black arrowin Fig. S2). The second instance that grew in the twinned direction was two bilayers thick (marked with black circle in Fig. S2). This double bilayer was extremely slow to grow and indicates that if due to some reason multiple layers are growing together then the growth of each layer is much slower than if just one layer was growing at a time (which is not surprising).

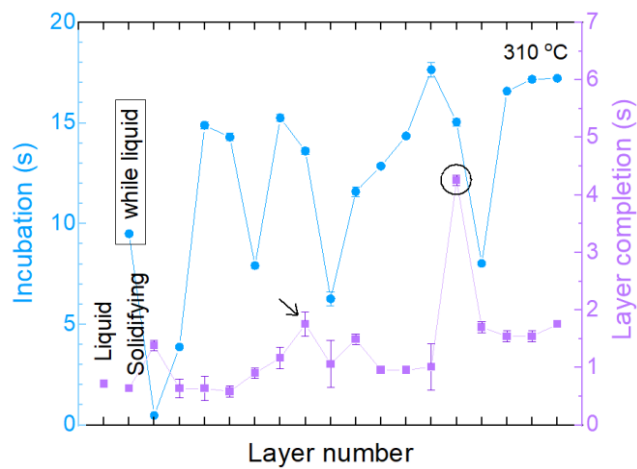

Fig. S2. Layer growth at 310 °C. Y-axis are in linear scale. The black arrow indicates a single layer growing as a twin. The black circle indicates a twinned double layer.

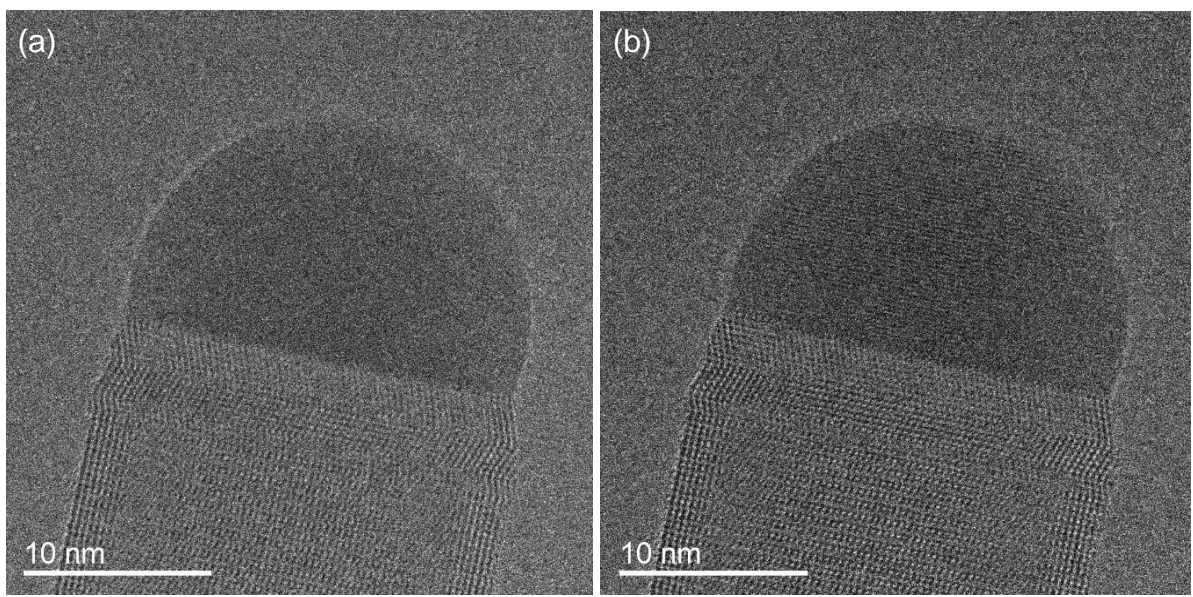

Fig. S3. (a) The video frame just before solidification. The very next frame in the video is given as (b). In (b) a crystalline lattice is clearly visible. (Exposure time of each image in the video was 0.1596s, corresponding to 6.27 fps.)

## 3) Diameter evolution

While growing in the zinc blende structure the catalyst size was changing. The diameter of the catalyst-nanowire interface at 320 °C (VLS) and 310 °C (VSS), along with the incubation and layer completion times, are shown in Fig. S4. The diameter is measured just before the beginning of that layer. Except at the transition stage due to solidification, at 310 °C the diameter of the solid catalyst kept increasing while the height consistently kept decreasing.

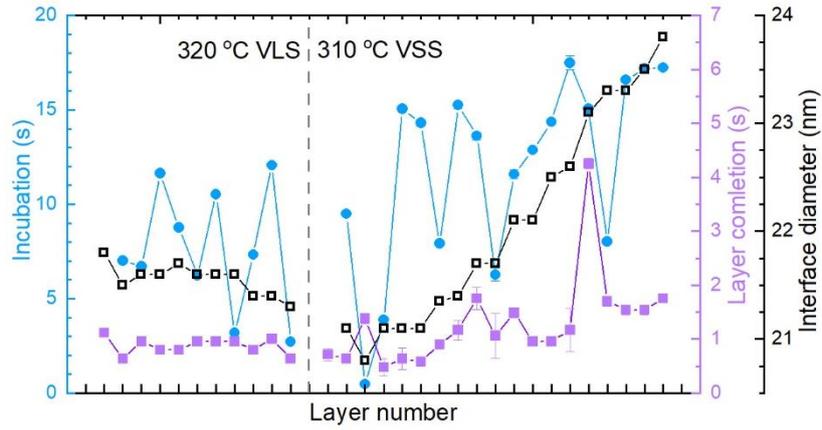

Fig. S4. The diamter of the catalyst-nanowire interface at 320 °C (VLS) and 310 °C (VSS except the first layer) plotted along with the layer growth times.

**4) Geometry based model for layer completion time**

We try to understand the increase of layer completion time with increasing interface diameter by a very crude geometric model. There could be other factors also related to the layer completion time, however, here we try to look at the geometry alone. Due to the negligible surface diffusivity, arsenic atoms reach the catalyst by direct impingement on the catalyst surface.[42] Thus the number of As atoms being adsorbed is proportional to the surface area of the catalyst exposed to the surface. Arsenic desorption is also proportional to the surface area.[49] Thus, the layer completion time would be inversely proportional to the catalyst-vapor surface area. Since the GaAs layer is formed at the catalyst-nanowire interface, the layer completion time varies proportionally to the interface area. For the VSS case the catalyst shape can be very crudely approximated to a cylinder with diameter D. (Note that the actual catalyst shape is not of a cylinder, a closer approximation would be that of a spherical or conical section placed above a cylinder and will involve many more parameters.) We assume that while the catalyst is flattening the catalyst volume ($V$) remains unchanged. According to this simplistic model the layer completion would vary as $\frac{1}{1+\frac{16\,V}{\pi\,D^3}}$, like shown below. The plot shows a trend similar to that in Fig. 4a of the manuscript.

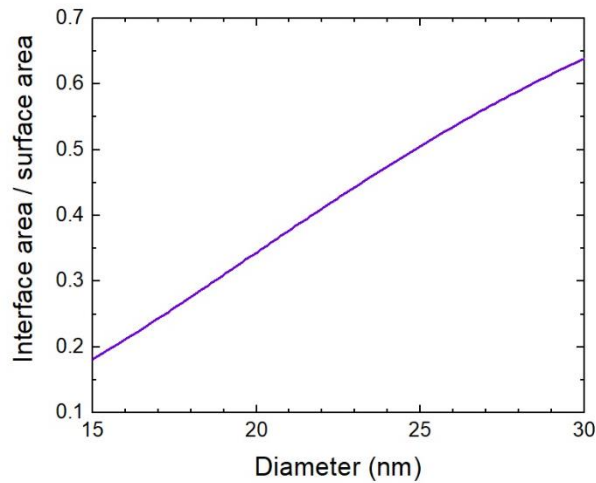

Fig. S5. NW-catalyst interface area divided by the catalyst surface area exposed to the ambient vapor plotted as a function of the cylinder diameter with a fixed catalyst volume of 3000 nm$^3$.

25